# Determining Reaction Pathways at Low Temperatures by Isotopic Substitution: The Case of BeD$^+$ + H$_2$O

Tiangang Yang[1,2*], Bin Zhao[3*], Gary K. Chen[1], Hua Guo[4], Wesley C. Campbell[1,5,6] and Eric R. Hudson[1,5,6]

[1] Department of Physics and Astronomy, University of California Los Angeles, Los Angeles, California 90095, USA

[2] Department of Chemistry, Southern University of Science and Technology, Shenzhen 518055, China

[3] Theoretische Chemie, Fakultät für Chemie, Universität Bielefeld, Universitätsstraße 25, D-33615 Bielefeld, Germany

[4] Department of Chemistry and Chemical Biology, University of New Mexico, Albuquerque, New Mexico 87131, USA

[5] UCLA Center for Quantum Science and Engineering, University of California – Los Angeles, Los Angeles, California 90095, USA

[6] UCLA Challenge Institute for Quantum Computation, University of California, Los Angeles, California 90095, USA

* E-mail: yangtg@sustech.edu.cn , bin.zhao@uni-bielefeld.de

**Abstract**

Trapped Be$^+$ ions are a leading platform for quantum information science [1], but reactions with background gas species, such as H$_2$ and H$_2$O, result in qubit loss. Our experiment reveals that the BeOH$^+$ ion is the final trapped ion species when both H$_2$ and H$_2$O exist in a vacuum system with cold, trapped Be$^+$. To understand the loss mechanism, low-temperature reactions between sympathetically cooled BeD$^+$ ions and H$_2$O molecules have been investigated using an integrated, laser-cooled Be$^+$ ion trap and high-resolution Time-of-Flight (TOF) mass spectrometer (MS) [2]. Among all the possible products, BeH$_2$O$^+$, H$_2$DO$^+$, BeOD$^+$, and BeOH$^+$, only the BeOH$^+$ molecular ion was observed experimentally, with the assumed co-product of HD. Theoretical analyses based on explicitly correlated restricted coupled cluster singles, doubles, and perturbative triples (RCCSD(T)-F12) method with the augmented correlation-consistent polarized triple zeta (AVTZ) basis set reveal that two intuitive direct abstraction product channels, Be + H$_2$DO$^+$ and D + BeH$_2$O$^+$, are not energetically accessible at the present reaction temperature (~150 K). Instead, a double displacement BeOH$^+$ + HD product channel is accessible due to a large exothermicity of 1.885 eV through a submerged barrier in the reaction pathway. While the BeOD$^+$ + H$_2$ product channel has a similar exothermicity, the reaction pathway is dynamically unfavourable, as suggested by a Sudden Vector Projection analysis. This work sheds light on the origin of the loss and contaminations of the laser-cooled Be$^+$ ions in quantum-information experiments.

**1. Introduction**

Trapped atomic ions such as Be$^+$, Mg$^+$, Ca$^+$, Ba$^+$, and Yb$^+$ are used to host qubits for quantum information processing and have been employed to construct fully programmable quantum computers [3, 4]. They have demonstrated the longest coherence times [5] and are prized for their extremely low single- and two-qubit gate error rates [1, 6, 7]. Among the various species in use, Be$^+$ is the lightest, providing the highest trap frequency which furnishes several advantages including reduced gate times. Though experiments are performed in ultrahigh vacuum, chemical reactions between Be$^+$ and the background gases (e.g. H$_2$ and H$_2$O) are the dominant processes for qubit loss [8, 9]. For instance, Be$^+$ is known to react with H$_2$ to produce BeH$^+$ when excited to the *p*-state by the 313 nm cooling laser [10]. In order to reverse BeH$^+$ to Be$^+$, photodissociation by 157 nm photons has been applied, but this did not recover all of the original Be$^+$ ions, and other impurities and the chemical reactions behind the creation of the remaining species, which may reveal similar rescue methods, are not well understood [11].

Thanks to an ion trap-integrated high-resolution time-of-flight mass spectrometer (TOF-MS) [2, 12-14], we are able to investigate chemical reactions of laser-cooled $Be^+$ ions with different background molecules in the trap. In our previous work, we determined that $Be^+$ ions react with $H_2O$ to produce $BeOH^+$. The reaction rate agrees with the capture theory when $Be^+$ is in the electronically-excited *p*-state, but a submegerd barrier in the *s*-state reaction significantly reduces the groud-state reaction rate [15, 16]. In this work, we report that the $BeH^+$ product from the $Be^++H_2 \rightarrow BeH^++H$ reaction can further react with $H_2O$ to produce $BeOH^+$. To better understand the reaction mechnism(s), isotope subsitution was employed to probe details about reaction dynamics and identify reaction pathways [17-24]. To this end, gaseous $D_2$ was introduced instead of $H_2$ to initially produce $BeD^+$ in the trap, which is then allowed to react with $H_2O$ and the charged products are analysed with the TOF-MS. Interestingly, among several possible products only a double displacement channel ($BeOH^++HD$) is observed experimentally. This observation is rationalized with *ab initio* electronic structure calculations of the reaction pathways, using explicitly correlated restricted coupled cluster singles, doubles, and perturbative triples (RCCSD(T)-F12) method and the augmented correlation-consistent polarized triple zeta (AVTZ) basis set.

## 2. Experiment

The apparatus employed here has been described in detail elsewhere [15, 16]. Briefly, $Be^+$ ions, produced from laser ablation of metallic Be, are trapped in a linear radio frequency Paul trap [14]. Laser cooling [25] is used to cool the translational motion of the ions, resulting in a Coulomb crystal of $Be^+$ ions, whose fluorescence is monitored by a camera in real time. A mixture of gaseous $D_2$ and $H_2O$ is then introduced via a room-temperature leak valve into the trapping region, where the gas reacts with the trapped ions. In order to produce higher amounts of $BeD^+$ for further reaction, $D_2$ is about 100x more prevalent than $H_2O$ in the mixture. The gaseous $D_2$ is measured from a residual gas analyzer (RGA), and the calibration of its fractionation has been described in our previous work [15, 16]. While the $H_2O$ density is too low for the RGA to detect accurately, its density is estimated later from the reaction rate. The ionic products remain in the trap, and are subsequently analyzed by an integrated TOF-MS [2, 12-14]. The 313 nm laser for cooling $Be^+$ ions allows manipulation of the $Be^+$ electronic quantum states; by tuning the frequency of this cooling laser, the fraction of ions in the $^2S_{1/2}$ and $^2P_{3/2}$ states can be precisely controlled [15].

The translational energy of the laser-cooled Be$^+$ ions is less than 1 K, while the translational and internal energy of D$_2$ and H$_2$O is assumed to be given by a thermal distribution at 300 K. The BeD$^+$ ions from the reaction of Be$^+$ + D$_2$ are sympathetically cooled by Be$^+$ and further reacted with room-temperature H$_2$O. The reaction temperature of BeD$^+$ + H$_2$O is roughly 150 K. Typical TOF traces (10 sample average) at reaction time t = 0 s and 60 s with 28% relative Be$^+$ $^2$P$_{3/2}$ state excitation ($P_P \approx 28\%$) and a full reaction curve are shown in Figure 1 in the left and right panels, respectively. The reaction time zero is determined from the fluorescence signal monitored by the camera in real time, which is also used to normalize the initial ion number for the TOF. At t = 0 s, only one clear peak ($m/z = 9$, Be$^+$) is shown in the TOF trace (red line). After 60 s, two more peaks arise at $m/z = 11, 26$, which indicates that BeD$^+$ and BeOH$^+$ are the two main products for Be$^+$ + D$_2$, H$_2$O reactions.

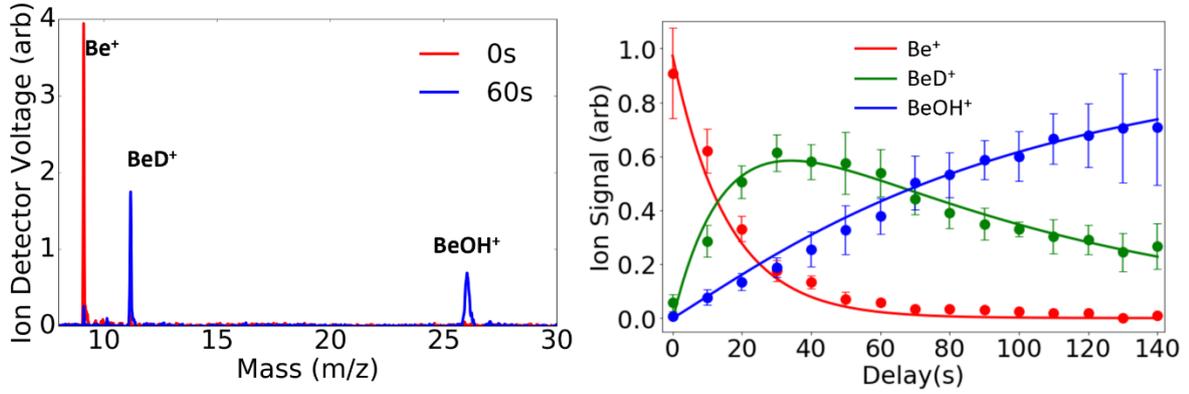

**Figure 1.** (Left) TOF signals (averaged over 10 trials) at reaction time t = 0 s and 60 s with $P_P \approx 28\%$. Three clear peaks at $m/z = 9, 11, 26$ are shown and interpreted as Be$^+$, BeD$^+$ and BeOH$^+$ ions, respectively. (Right) The temporal evolution of Be$^+$, BeD$^+$ and BeOH$^+$ in the trap as a function of reaction time as well as the solutions of differential equations fitted to the kinetics data with $P_P \approx 28\%$.

The reactions of interest are:

Be$^+$ + H$_2$O → BeOH$^+$ + H  (1)

Be$^+$ + D$_2$ → BeD$^+$ + D  (2)

BeD$^+$ + H$_2$O → BeOH$^+$ + HD  (3)

BeD$^+$ + H$_2$O → BeOD$^+$ + H$_2$  (4)

Thus, the kinetics of the reagents and products are found from:

$d[\text{Be}^+]/dt = -(k_1\rho_1 + k_2\rho_2)[\text{Be}^+](t)$  (5)

$d[\text{BeD}^+]/dt = k_2\rho_2[\text{Be}^+](t) - k_3\rho_1[\text{BeD}^+](t)$  (6)

$$d[BeOH^+]/dt = k_1\rho_1[Be^+](t) + k_3\rho_1[BeD^+](t) \qquad (7)$$

where $k_i$ is the reaction rate coefficient for reactions $i = (1) – (3)$, reaction (4) is not included in the fitting because no evidence of $BeOD^+$ has been shown in the TOF signals (Figure 1 (left)). $k_1, k_2$ are set as known in the fitting from the previous measurements [10, 15]. $\rho_1, \rho_2$ are the density of $H_2O$ and $D_2$, respectively, where $\rho_2$ is measured to be $\sim 1.5 \times 10^8$ molec./cm$^3$ by the RGA, $\rho_1$ and $k_3$ are estimated from the fitting. The temporal evolution of $Be^+$, $BeD^+$ and $BeOH^+$ in the trap as a function of reaction time are shown in Figure 1 (right), as well as the solutions of differential equations fitted to the kinetics data. The reaction rate coefficient of $k_3$ has been measured to be $(3.5 \pm 2) \times 10^{-9}$ cm$^3$/s, while $\rho_1$ is estimated at $\sim 3 \times 10^6$ molec./cm$^3$.

It is interesting to note that although reactions (3) and (4) appear to formally lead to the same products, they represent very different pathways. Reaction (3) preseves one O-H bond in the $H_2O$ moiety, which can be achieved by breaking the Be-D and H-O bonds and forming a new H-D bond. On the other hand, reaction (4) would require O to insert into the Be-D bond, which involves the breaking of both O-H bonds. Such a drastic difference cannot be distinguished if all three hydrogens are the same isotope, but the difference is reavealed through isotope tagging.

## 3. Theory

To investigate the dynamics behind the experimental observation in Figure 1, we theoretically explore the critical stationary points in the $BeD^+(^1\Sigma)+H_2O$ reaction pathways using the explicitly correlated restricted coupled cluster singles, doubles, and perturbative triples (RCCSD(T)-F12) method [26] with the augmented correlation-consistent polarized triple zeta (AVTZ) basis set [27]. The results are shown in Figure 2, and the structures of the corresponding stationary points are given in Table S1. It should be noted that at the experimental conditions, the reactants do not have enough overall energy to overcome any significant reaction barrier. Indeed, the reaction pathways shown in Figure 2 only feature submergered barriers.

Two intuitive pathways featuring the abstraction of either $D^+$ or $Be^+$ from $BeD^+$ by $H_2O$ lead to the $Be+H_2DO^+$ and $D+BeH_2O^+$ channels, respectively. The two pathways are initiated with barrierless attractive interaction between the cationic $BeD^+$ and the lone electron pair on the oxygen atom of the dipolar $H_2O$, as shown in Figure 3. In the $Be+H_2DO^+$ channel, there exist two shallow wells (MIN1 and MIN2) corresponding to the $BeD^+$-$H_2O$ and the $Be$-$H_2DO^+$ complexes, which are separated by a moderate saddle point (SP1). The classical energy of the $Be+H_2DO^+$ product channel is similar to that of the reactant. However, when zero-point energies of the reactants and products are considered this product channel lies 0.218 eV above the reactant. As a result, it is inaccessible under the experimental conditions. The second channel, $D+BeH_2O^+$, involves a deep well (MIN3) formed with the Be-end pointing to the water oxygen, as the Be end has a more positive charge. The D+BeH$_2$O$^+$ products are realized by breaking the Be-D bond in the DBe$^+$-OH$_2$ complex. This product asymptote is also endoergic by 0.423 eV and unlikely to occur at the present experimental reaction temperature (~ 0.013 eV), which is consistent with the absence of this products in the experiment.

The third pathway leads to the BeOH$^+$+HD product channel, which lies 1.885 eV below the reactant energy. This pathway is also initiated by the barrierless formation of the ion-dipole DBe$^+$-OH$_2$ complex (MIN3), followed by a submerged saddle point (SP2) to another well (MIN4), namely the HD-BeOH$^+$ complex. The SP2 features a double displacement process by forming a Be-O bond and transfering one H atom from H$_2$O to the D atom in BeD$^+$ reactant. This double displacement reaction is difficult, evidenced by a large barrier (1.883 eV at SP2) from the DBe$^+$-OH$_2$ well (MIN3), due apparently to the breaking of two bonds (i.e., O-H and Be-D bonds). However, this difficult bond rearrangement is feasible thanks to the 1.700 eV of energy available at this submergerd barrier (SP2). The reaction completes by the escape of the neutral, low-mass HD product from the trap. The production of the HD product, instead of the H$_2$ product absent in the experiment that would only be formed from H$_2$O, suggests that the two hydrogens have to come from the two reactants,. This observation demonstrates the power of the isotope substitution in identifying the reaction pathway.

Perhaps the most surprising result of this study is that though the isotopologue BeOD$^+$+H$_2$ product channel is also exothermic, this channel is not observed in the experiment. This product channel can be reached by either insertion of O into the BeD$^+$ or via a triple displacement channel (shown in Figure S1), which exchanges the hydrogen moieties connected with O and Be. The first pathway, insertion of O into the BeD$^+$, is unlikely because the strong ion-dipole attraction between the reactants leads to either the BeD$^+$-OH$_2$ (MIN1) or the DBe$^+$-OH$_2$ (MIN3) wells, as shown in Figure 3. Furthermore, the insertion mechanism requires the simultaneous breaking of the two O-H bonds, in addition to the Be-D bond cleavage, which are extremely difficult. The second possible mechanism, namely the triple displacement channel (shown in Figure S1), would have to exchange the three hydrogen moieties connected with O and Be. From the MIN4 complex, the triple displacement channel requires two additional transition states, which have the same geometry as SP2 but different permutations of the hydrogens. In the first transition state, the D atom is transferred from Be to O, forming the HBe$^+$-OHD complex (MIN3'). This is followed by the second transition state, where the remaining H on O is transferred to Be, leading to the H$_2$-BeOD$^+$ complex (MIN4'), which can then dissociate to form the H$_2$ + BeOD$^+$ product.

The absence of this BeOD$^+$+H$_2$ product channel in the experiment can be rationalized by the Sudden Vector Projection (SVP) model [28], which attributes the product energy disposal by the projection of a product mode onto the reaction coordinate at the transition state (SP2). The SVP model is based on the premise that the projection ($\eta \in [0,1]$) is a proxy for the coupling strength of the product mode with the reaction coordinate, thus dictating the amount of the enregy flow into the corresponding product mode as the system departs from the transition state to the product channel. This model has been tested in a large number of reactions and its predictions are quite reliable [29].

SVP values (see Table S2 and Figure S2) for SP2 suggest that the center-of-mass separation mode between the two products has a large SVP value ($\eta = 0.50$) with SP2, suggesting that a large portion of the energy release is expected to be partitioned into the translational energy between HD and BeOH$^+$, leading to the dissociation. This SVP prediction suggests that MIN4 is probably a short-lived species formed transiently on the way to the products. This is further aided by the fact that SP2 has a much more restricted geometry (a small volume in the phase space) than the product channel (much larger phase-space volume), as the two products can take arbitray geometries and internal excitations. So even if the MIN4 complex were sufficiently long lived, the chances for dissociating into the product channel would be much larger than those for going back to MIN3 via SP2. The difficulties for the triple displacement pathway are further compounded by the need to pass two more SP2-like transition states, as dipected by Figure S1. Thus, these dynamical factors, namely the large SVP value for the dissociation coordiante and favorable entrope of the product channel, , argue against the the triple displacement channel, explaining the absence of the H$_2$ + BeO$^+$ products.

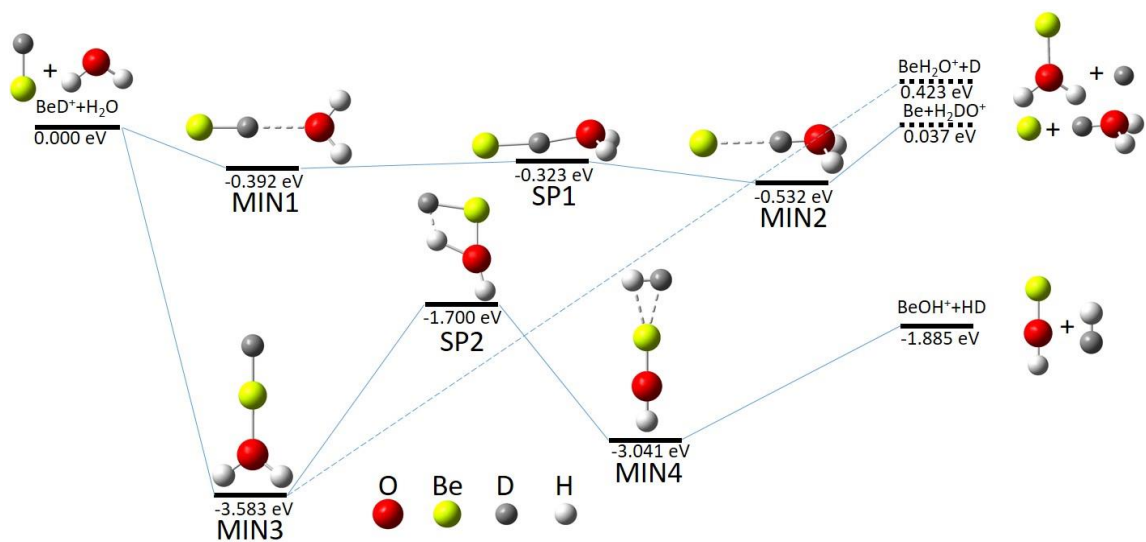

**Figure 2**. Energetics and structures of stationary points for two BeD$^+$ + H$_2$O reaction pathways. The stationary points were calculated at the RCCSD(T)-F12/AVTZ level of theory.

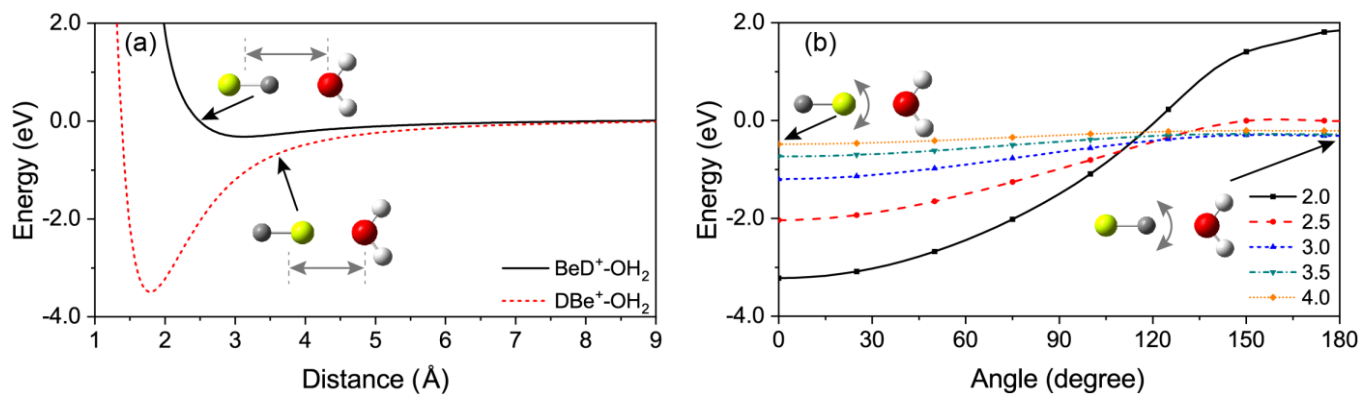

**Figure 3**. 1D cuts of the potential energy surface of the $BeD^+ + H_2O$ reaction. (Left) Energy with respect to the distance between the O atom and the center of mass (COM) of $BeD^+$ at the $BeD^+$-$OH_2$ and $DBe^+$-$OH_2$ configurations. (Right) Energy with respect to the orientation of the $BeD^+$ reactant at five different distrance between $BeD^+$ and $H_2O$. In the plots, the atoms lie in a common plane, and the $BeD^+$ and $H_2O$ reactants were fixed at the equilibrium geometries. The COM of $BeD^+$ was restricted to be along the bisector of the $H_2O$ molecule.

To summarize, reactions between cold $BeD^+$ and warm $H_2O$ have been investigated using TOF-MS. This reaction readily produces $BeOH^+$. However, the $BeOD^+$ product was never observed, despite its similar energy to $BeOH^+$. The reaction mechanism was analyzed using a high-level *ab initio* method and the results confirm the observation that only one reaction path, leading to $BeOH^+ + HD$, is realized under experimental conditions. Deuteration of the $BeH^+$ reactant offered a unique opportunity to identify the reaction pathway in this important ion-molecule reaction, which represents a potentially important loss-channel for $Be^+$ qubits if both $H_2$ and $H_2O$ residual gases exists in the vacuum chamber. While previous work has shown that $Be^+$ qubits can be recovered from $BeH^+$ with photodissociation at 157 nm, recovery of $Be^+$ from the $BeOH^+$ molecular ion requires photodissociation between approximately 223 nm and 232 nm.

**Acknowledgements**

The authors thank Arthur Suits for helpful discussions. The UCLA authors thank Michael Heaven and Hao Wu; BZ thanks Uwe Manthe and Wolfgang Eisfeld for helpful discussions. Funding: This work was supported by the Air Force Office of Scientific Research grant nos. FA9550-16-1-0018, FA9550-20-1-0323 and FA9550-18-1-0413 (HG). HG also thanks the Alexander von Humboldt Foundaiton for a Humboldt Research Award. T.Y. thanks support from Guangdong Innovative &

Entrepreneurial Research Team Program (Grants 2019ZT08L455) and the National Natural Science Foundation of China (Grant 22003023, NSFC Center for Chemical Dynamics).

Supporting information:

Table S1 Energies and geometries of the stationary points for the BeD$^+$+H$_2$O→BeOH$^+$+HD reaction. The results are obtained by ab initio calculations using explicitly correlated restricted coupled cluster singles, doubles, and perturbative triples (RCCSD(T)-F12) method with the augmented correlation-consistent polarized triple zeta (AVTZ) basis set. (The reactive hydrogen atom in H$_2$O molecule is denoted as H2, and the other one as H3.) Energy, bond length, and angle are in units of eV, Angstrom, and degree, respectively.

|  | E | $r_{OH2}$ | $r_{OH3}$ | $r_{BeD}$ | $r_{BeO}$ | $\theta_{H2OH3}$ | $\theta_{BeOH2}$ | $\theta_{DBeO}$ |
|---|---|---|---|---|---|---|---|---|
| BeD$^+$+H$_2$O | 0 | 0.9586 | 0.9586 | 1.3087 |  | 104.44 |  |  |
| MIN1 | -0.392 | 0.9608 | 0.9608 | 1.3438 | 3.3511 | 104.62 | 127.62 | 0.15 |
| SP1 | -0.323 | 0.9660 | 0.9660 | 1.5069 | 2.9619 | 106.79 | 116.72 | 3.55 |
| MIN2 | -0.532 | 0.9736 | 0.9736 | 2.0651 | 3.1024 | 110.35 | 111.19 | 0.80 |
| MIN3 | -3.583 | 0.9720 | 0.9720 | 1.3087 | 1.5562 | 109.15 | 125.42 | 180.00 |
| SP2 | -1.700 | 1.2828 | 0.9470 | 1.4222 | 1.4147 | 134.12 | 64.81 | 97.42 |
|  | E | $r_{OH3}$ | $r_{BeO}$ | $r_{BeD}$ | $r_{DH2}$ | $\theta_{H2BeD}$ | $\theta_{BeOH3}$ | $\theta_{DBeO}$ |
| MIN4 | -3.041 | 0.9539 | 1.3369 | 1.6220 | 0.7776 | 27.74 | 180.00 | 166.13 |
| BeOH$^+$+HD | -1.885 | 0.9390 | 1.3122 |  | 0.7579 |  | 180.00 |  |

Table S2 Projections of the HD+BeOH$^+$ product translational, vibrational, and rotational modes onto the reaction coordinate at the saddle point SP2, along with the vibrational frequencies of the HD and BeOH$^+$ products.

|  |  | Vibrational frequency (cm$^{-1}$) | SVP values |
|---|---|---|---|
|  | Translational mode |  | 0.50 |
| HD | vibration | 3826.28 | 0.89 |
|  | rotation |  | 0.38 |
| BeOH$^+$ | in-plane bend | 391.29 | 0.07 |
|  | out-of-plane bend | 391.29 | 0.00 |
|  | OBe stretch | 1566.35 | 0.06 |
|  | OH stretch | 3992.39 | 0.03 |
|  | rotation |  | 0.03 |

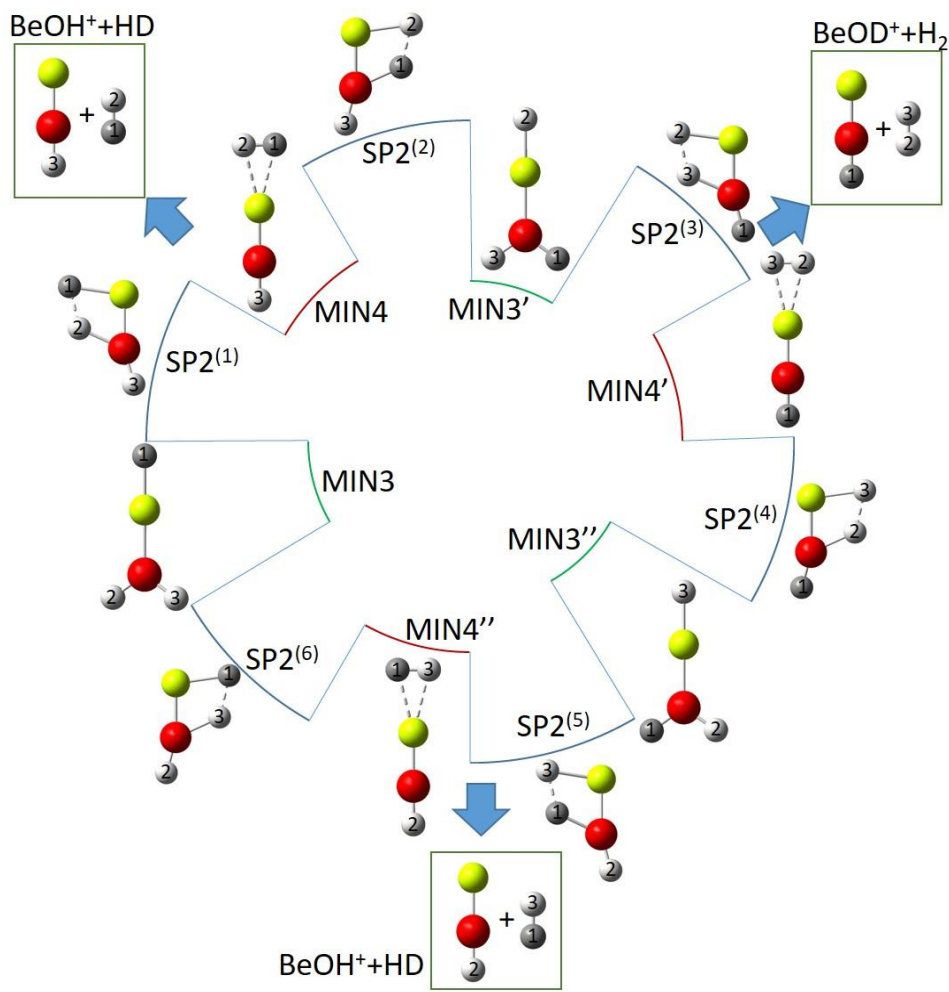

Figure S1. Schemicatic of the triple displacement mechamism, which allows the permuation of all the three Hydrogen species and leads to three isotopologues product channels.

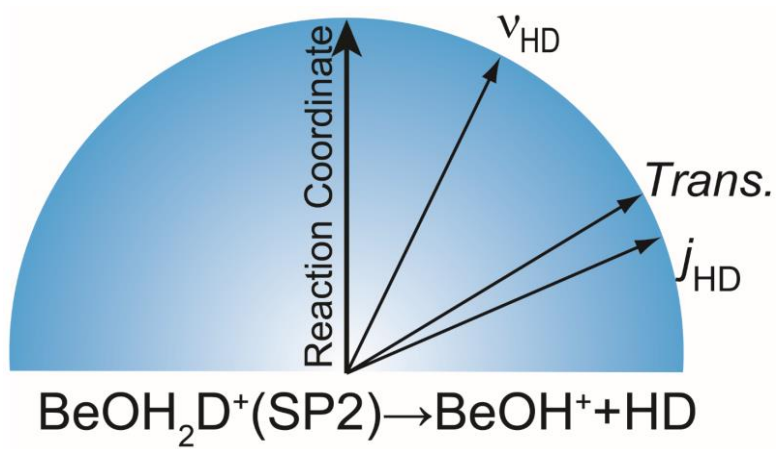

Figure S2 . Alignment of the translational, HD vibrational, and HD rotaional vectors with the reaction coordinate at the saddle point SP2. The ones for the BeOH$^+$ coproduct are almost perpendicular to the reaction coordinate. Please refer to Table S2 for more details.